\documentclass[graybox]{svmult}

\usepackage{geometry}
\usepackage{ebgaramond,ebgaramond-maths,makeidx,graphicx,multicol}
\usepackage[bottom]{footmisc}
\graphicspath{{Figures/}}
\makeindex

\begin{document}

\title*{A Map of Approaches to Temporal Networks}

\author{Petter Holme and Jari Saram\"aki}

\institute{Petter Holme \at Institute of Innovative Research, Tokyo Institute of Technology, Japan \email{holme@cns.pi.titech.ac.jp}
\and Jari Saram\"aki \at Department of Computer Science, Aalto University, Finland \email{jari.saramaki@aalto.fi}}

\maketitle

\abstract{The study of temporal networks is motivated by the simple and important observation that just as network structure can affect dynamics, so can structure in time. Just as network topology can teach us about the system in question, so can its temporal characteristics. In many cases, leaving out either one of these components would lead to an incomplete understanding of the system or poor predictions. We argue that including time into network modeling inevitably leads researchers away from the trodden paths of network science. Temporal network theory requires something different---new methods, new concepts, new questions---compared to static networks. In this introductory chapter, we overview the ideas that the field of temporal networks has brought forward in the last decade. We also place the contributions to the current volume on this map of temporal-network approaches.}


\section{Overview}
\label{sec:overview}

\begin{figure}
\begin{center}
  \includegraphics[width=0.90\textwidth]{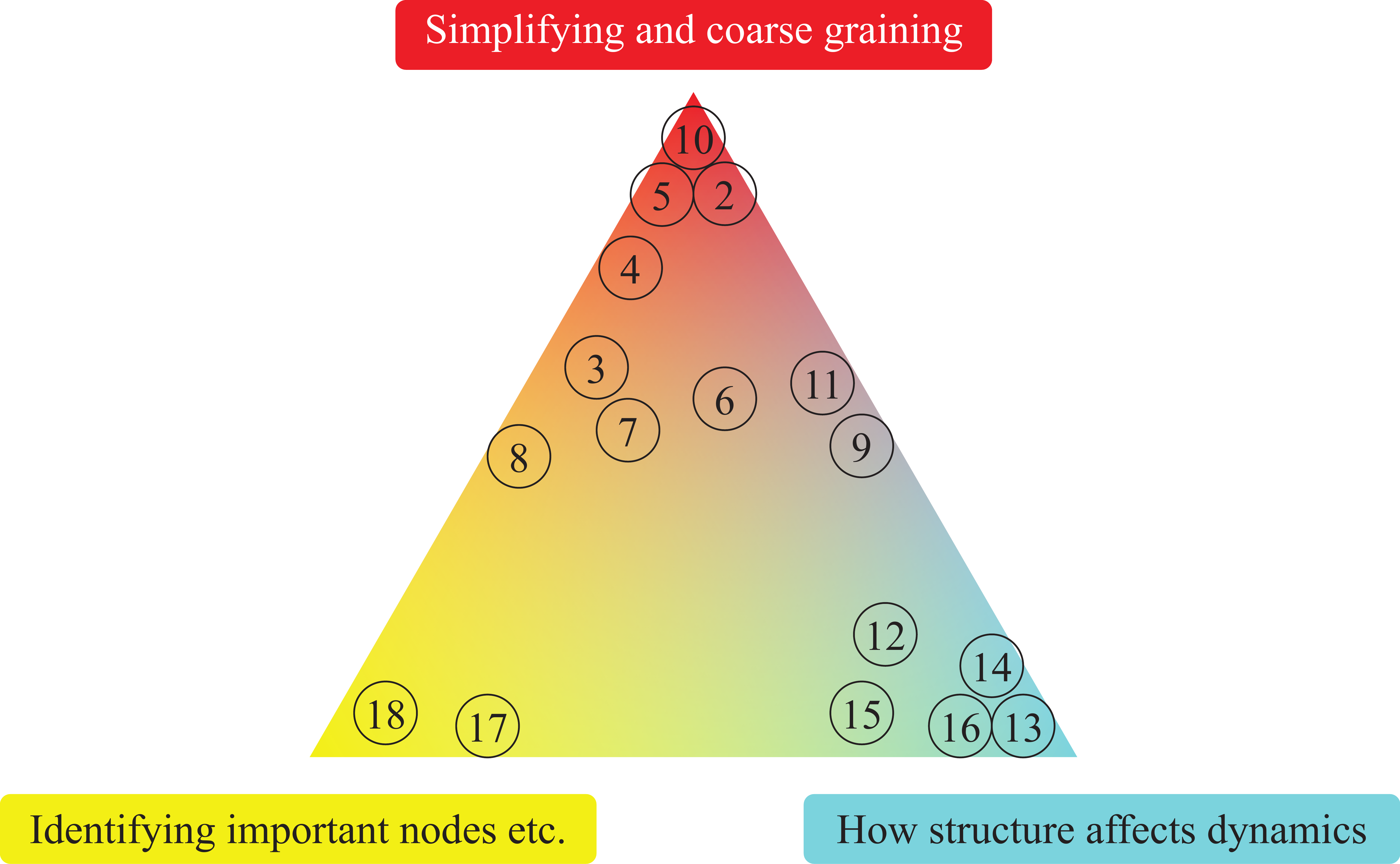}  
\end{center}
\caption{A schematic map of the chapters of this book, positioned with respect to the three main research themes within the study of temporal networks.}
\label{fig:triangle}
\end{figure}

If we want to make sense of large, complicated systems via the data they leave behind, we need to simplify them systematically. Such simplifications typically need to be very drastic. A common first step is to represent the system as a network that only stores information on which units are connected to which other units. To investigate the World Wide Web with this approach, one would neglect the content, the owner, the time of creation, and the number of downloads of a web page. Instead, one would only consider individual webpages and how they are linked together. The second step is to apply network science methods to find important nodes or clusters of nodes with some special role or function or study how the network's wiring controls some dynamical system. The fundamental idea of this book is that one can learn more about a system if one does not, at the first step of simplification, discard information about when things happen. Consequently, one needs to modify the second step and develop a science of temporal networks that exploits this additional information.

The fundamental idea of retaining information about time is evidently not a hard one to get. Temporal networks have been invented and reinvented many times. Researchers have proposed many mathematical and computational frameworks---some equivalent, some not. This is probably inevitable for such an extraordinarily interdisciplinary field of science---temporal networks have been applied to neuroscience, transportation problems, social theory~\cite{brudner1997class}, control theory~\cite{LiCornelius2017Science}, ecology~\cite{Ushio2018Nature}, and many more areas. The many existing frameworks could be frustrating for a newcomer to temporal networks. Part of our idea with this book was to showcase this diversity; see Chapters 2, 3, and 15 for very different ways of thinking about networks in time.

Even if you encounter a problem where both the network and the temporal aspects should play a role, there is no general recipe to follow. This introductory chapter aims to provide a rough map of the field--what types of questions researchers have been interested in and what results are out there. We will also try to place the subsequent chapters in their correct locations on this map (Fig.~\ref{fig:triangle}). This chapter is not a catalog of techniques or an introduction to a comprehensive and self-consistent theory. For readers interested in that, our review papers~\cite{HolmeSaramaki2012PhysRep,Holme2015EurPhysJB} the book by Masuda and Lambiotte~\cite{Masuda2016book} or by Batagelj \textit{et al.}~\cite{vlado2014} will be a better read.

\section{Temporal network data}
This section discusses the many subtleties about how to represent a system as a temporal network in a meaningful way.

\subsection{Events}
\label{sec:events}

The fundamental building blocks of temporal networks are events (or contacts, links, or dynamic links). These represent units of interaction between a pair of nodes at specified times. Often, they take the form of triples $(i,j,t)$ showing that nodes $i$ and $j$ are in contact at time $t$. Sometimes the time can be an interval, rather than just a moment.

As we will see throughout this book, temporal-network modeling is far from a straightforward generalization of static networks---often, it is fundamentally different. As a first example, we note that events are not always a straightforward generalization of the static networks' links. Take e-mail communication as an example. In static network modeling, one typically assumes that the links (between people that have exchanged e-mail) indicate social relationships. These links can be viewed as the underlying infrastructure for \textit{e.g.}\ information spreading since people who know each other exchange information. The links are there not only for one e-mail to be sent but represent persistent opportunities for spreading events for the duration of the relationship. In contrast, an event in a temporal e-mail network is simply one e-mail being sent. This usually happens for the explicit purpose of spreading information.
But there are also systems other than e-mail communication where events are more like links of static networks. Consider, for example, transportation systems, where the bus, train, or flight connections are really opportunities to travel that happen whether a certain person needs to use them or not. As we will see, different approaches treat these two interpretations of events differently.

\subsection{Boundaries}

In the natural sciences, we can sometimes model time as a dimension, if not exactly like space, then at least similar to it. For temporal networks, the binary connections and the time are more fundamentally different concepts. The simplest way of seeing this is to consider the network's boundaries (between what is contained in a data set and what is not). Regarding time, a temporal network data set almost always covers a time interval, and the interval is the same for all nodes. The structural boundaries of the network dimension are usually less controlled. Like cohort studies in the social sciences, one would like to select nodes that are as tight-knit as possible, typically defined by common features. For example, Ref.~\cite{Stopczynski} is based on data from voluntary participants among the freshmen of a university---better than a random group of people, but worse than the complete group of freshmen.

Boundaries become a problem when one wants to control the size of a data set. If a temporal network is too large to handle or one wants to understand the effects of size, how should one reduce the network's size without changing its structure? One could reduce the number of nodes by random sub-sampling or truncate the temporal data. However, both these approaches would introduce biases. While there are ways of correcting some of those~\cite{PhysRevE.92.052813}, it is hard to avoid problems. For example, if one truncates the data, there might not be enough time for a spreading process to saturate before the sampling interval is over. If one deletes nodes or events, one introduces other biases. The proper way of resampling a temporal network must simultaneously vary the number of nodes and the sampling duration, but exactly how is still an open question.

\begin{figure}
\sidecaption
\includegraphics[width=0.55\textwidth]{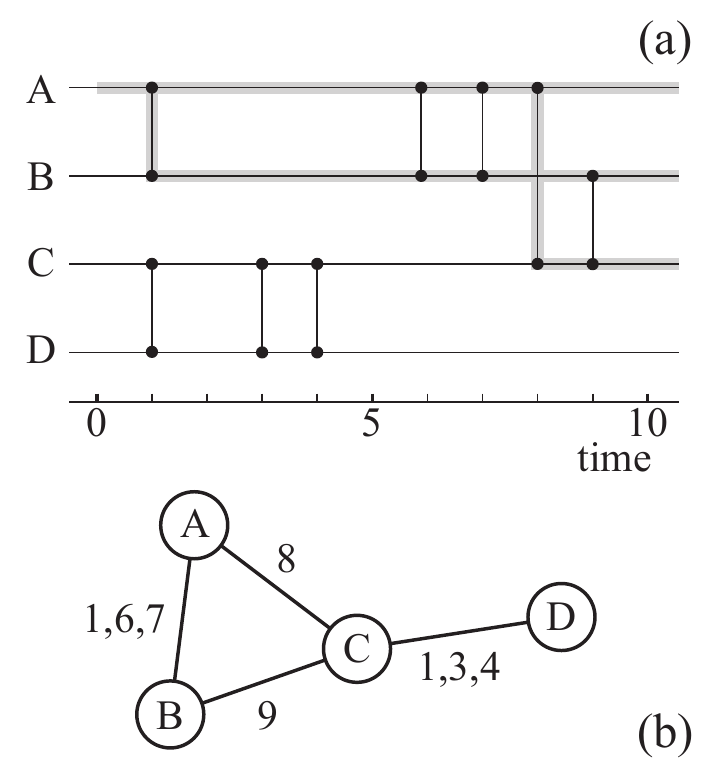}
\caption{An illustration of two ways to visualize small temporal networks that can be convenient for reasoning about measures and methods. Panel (a) shows a time-line graph where an epidemic outbreak starting at node A is indicated by grey lines. In almost all cases, paths between nodes (that follow events) in temporal networks need to go forward in time (to the right in the plot). (b) shows the same data but plotted projected onto a static graph. The latter visualization highlights the underlying static network structure at the expense of the temporal information. The former, the time-line plot, can capture many temporal structures but is inconvenient for network structures.}
\label{fig:connectivity}
\end{figure}

\subsection{Connectivity}

It is fundamental to network modeling that being indirectly connected through a path is relevant to dynamic processes. This is true for temporal networks as well, but the connections have to happen along time-respecting paths of contacts (with strictly increasing timestamps). Indirect connections through time-respecting paths are not transitive (see Fig.~\ref{fig:connectivity})---even if one can get from A to B and B to C, it might still be impossible to go from A to C because one would arrive at B too late for a further connection to be possible. Contrary to this, all static networks, directed networks included, are transitive.

\begin{figure}
\sidecaption
\includegraphics[width=0.60\textwidth]{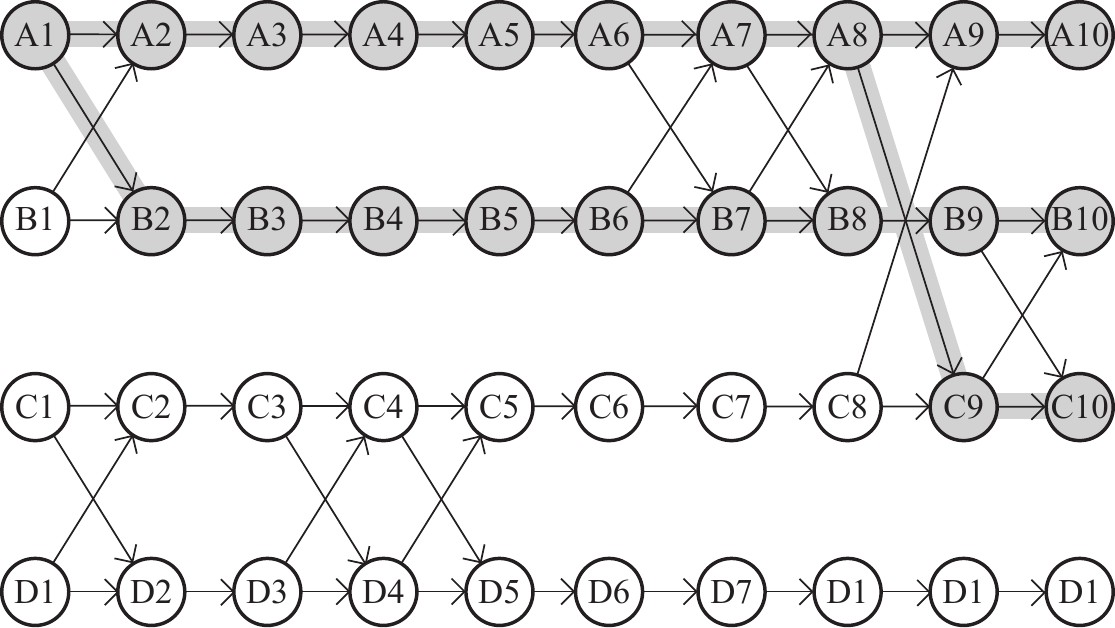}
\caption{A time-node representation of the data in Fig.~\ref{fig:connectivity}. This is a (directed, acyclic) static graph containing the same information as in Fig.~\ref{fig:connectivity} but the meaning of nodes and edges is different.}
\label{fig:timenode}
\end{figure}

Another important difference to static networks is that connectivity is temporal: even if there is a path from A to B now, whether direct or indirect, there might be none a second later. Therefore, the statement ``A is connected to B'' is not necessarily even meaningful unless the time (interval) of this connection is specified. 
The above issues mean that one can never reduce a temporal network into a static one without losing information or changing the meaning of the nodes
(\textit{cf.}\ Fig.~\ref{fig:timenode}).

Since many static network tools are based on paths and distances, researchers have sought to generalize these concepts to static networks. Once again, the addition of a temporal dimension makes this task much more complicated. The most common generalization of distance is \textit{latency} (\textit{temporal distance})~\cite{Lamport1978CommACM}---the time it would take to reach $j$ from $i$ starting at time $t$ and following only time-respecting paths. For a longer discussion about paths and connectivity, see Refs.~\cite{Masuda2016book,HolmeSaramaki2012PhysRep,Holme2015EurPhysJB}.

\section{Simplifying and coarse-graining temporal networks}

Even if representing data as a temporal network means that information has to be discarded for simplification, this is often not enough to understand the large-scale organization of the system. There are many ideas on how to simplify further a temporal network that we will discuss in this section.

\subsection{Projections to static networks}

Perhaps the most obvious way of simplifying a temporal network is to turn it into a static network. In fact, many classic examples of static networks like citation networks or affiliation studies (like the ``Southern Women'' study of 1941~\cite{southern_women}) have temporal link information. Still, the time is ignored by simply studying the network of all aggregated contacts or separate ``snapshot'' graphs representing different times.

If one,  from time-stamped data, constructs a binary static network where nodes are only linked or not, it is obvious that a lot of information is lost. A better option is to include information on the number or frequency of connections between pairs of nodes, leading to weighted networks. In this case, the link weights can provide important insights into the network structure (see, \emph{e.g.}, \cite{Barrat2004,Onnela_PNAS_2007}). However, including links between all nodes that have been in contact can, in some cases, result in a very dense network. In this case, one can threshold the network, discarding the weakest links, or extract the backbone of the network~\cite{backbone}. 

The above weighted-network approach is not really temporal because if one manipulates the times of the contacts, the outcome will remain the same. The simplest static networks that truly encode some temporal effects are \textit{reachability graphs}. These graphs have a directed edge $(i,j)$ if one can reach $j$ from $i$ via a time-respecting path.

Another way of creating sparser static networks than thresholding weighted graphs is to aggregate contacts within a time window~\cite{window}. While the thresholded graphs contain information about the most common contacts in the whole sampling interval, time-window graphs emphasize shorter time-scales, and their sequence captures at least part of the network dynamics. Indeed, tuning the time windows' duration can be a way to understand the organization of the data~\cite{sune_fundamental}. Yet a similar idea is to construct networks where links represent ongoing relationships~\cite{ongoing}---pairs of nodes that, at some point in time, have had contacts before and will have them again.

One more elaborate way of reducing temporal networks to static ones is the extraction of backbones, specifically concerning spreading processes on temporal networks~\cite{xiuxiu} and Chapter 11. By this approach, links in the resultant network correspond to node pairs that are likely to infect each other in an epidemic outbreak.

As mentioned above, these approaches can never retain all temporal features of the original data. Nevertheless, analyzing temporal networks by making them static is rather attractive because there are a plethora of methods for static-network analysis. 
One way of circumventing information loss is to use more elaborate mappings, where temporal networks are mapped onto static network structures whose nodes and links represent something else than the original network's nodes and links. One example is temporal event graphs, whose nodes correspond to the original network's events (see \cite{mellor_temporal_2018,kivela_mapping_2018} and Chapter 6 of this book). 

One common approach that can also be interpreted as static-network projection is to use multilayer networks, as in Chapter 17 of this book: time is sliced into consequent intervals, and the layers of a multilayer network correspond to networks aggregated for each interval. Once the layers are coupled (\textit{e.g.}\ with a directed link from a node to its future self), one can then apply (static) multilayer network methods to the system. Importantly, the layers in such a projection are ordered by time.

Finally, we can also project temporal network data to \textit{higher-order network models} that retains some information of the flows over the network. Chapter 4 discusses such approaches.

\begin{figure}
\includegraphics[width=\textwidth]{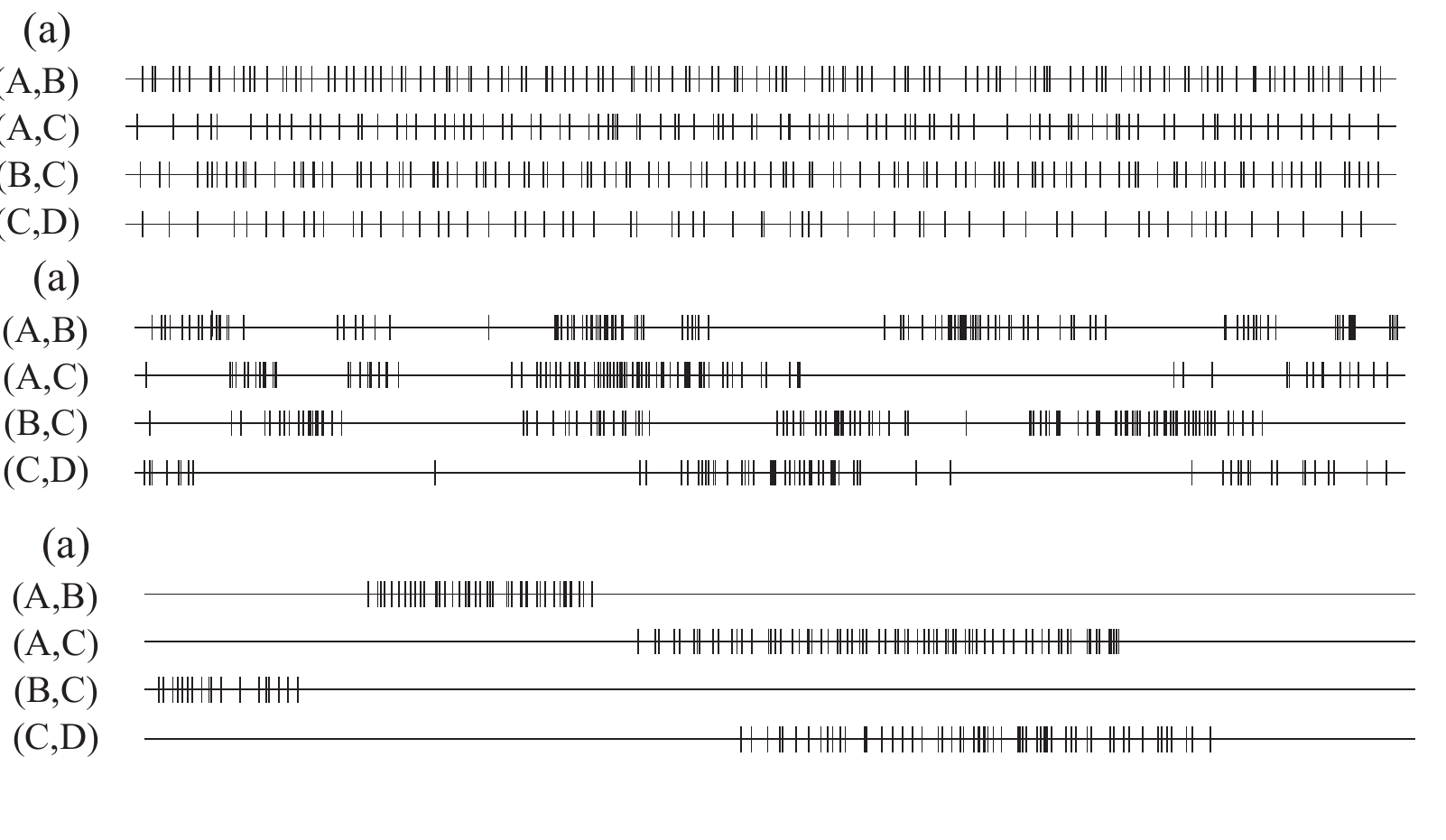}
\caption{Three scenarios of temporal edge structure. The figures show the time-lines of contacts along edges in a four-node graph. Panel (a) shows a scenario with narrow distributed inter-event times; (b) shows a bursty edge dynamics; (c) shows a scenario with a turnover of edges---where the time between the beginning of the sampling to the first contact, or from the last contact to the end of the sampling, is too long to be explained by the interevent time distributions.} 
\label{fig:temporal_pictures}
\end{figure}

\subsection{Separating the dynamics of contacts, links and nodes}

Instead of reducing temporal network data to static networks, one can retain some but not all of the temporal features. One example is the statistics of times between contacts. It was early recognized that often, the times between events, both for nodes and links, have heavy-tailed distributions~\cite{ongoing,johansen} (they are \textit{bursty}~\cite{barabasi_origin_2005,karsai:book}). Subsequent studies~(\emph{e.g.}\ Ref.~\cite{Karsai2011PhysRevE,MiritelloMoro}) found that this burstiness of inter-event times slows down spreading processes: simulated spreading that takes place on bursty networks is slower than it is on networks where the burstiness has been artificially removed. However, the result is the same when the heavy-tailed inter-event times are part of the dynamical process itself: when a spreading process with power-law distributed waiting times is placed on a static network~\cite{Min2011PhysRevE}, it is slow too. This is related to how events are interpreted (see \ref{sec:events} and Chapter 12): are they separated from the process and just passive conduits for it, as in spreading on top of bursty event sequences, or are the events actively generated by the process, as one could interpret the combination of spreading with broad waiting times and a static network?  Figs.~\ref{fig:temporal_pictures}(a) and (b) illustrate homogeneous and heterogeneous (bursty) link dynamics on top of a static network.  See also Chapter 9 that goes deeper into this issue.
Note that under some conditions, burstiness may also speed up spreading~\cite{RochaSimulated,Horvath2014}.

Another way of simplifying temporal networks is to ignore contact dynamics and think of links as present between the first and last observation of a contact in the data and ignore the precise timing of contacts~\cite{Holme2014SciRep}. Compared to simplifying the system as bursty dynamics on top of static networks, this picture emphasizes longer time structures such as the general growth and decline of activity in the data. Fig.~\ref{fig:temporal_pictures}(c) illustrates a data set that is well-modeled by links appearing and disappearing, disregarding the interevent time statistics.

\subsection{Mesoscopic structures}

In science, in general, ``mesoscopic'' refers to the scales between macroscopic and microscopic. In network science, this would mean structures larger than nodes but smaller than the entire network, and indeed,  the term is often used in the context of grouping nodes into classes based on how they are connected to each other and the rest of the network. The primary example of mesoscopic structures is community structure---that some networks have clear groups that are strongly connected within and weakly connected between each other~\cite{Schaub2017}. 

Most community detection methods in static networks divide the network so that every node belongs to one group only (Chapter 10). The straightforward extension of this idea to temporal networks would be to let nodes belong to different groups at different times, but only to a single group at each point in time~\cite{rossetti}. This is also the most common assumption in the literature, see \textit{e.g.}\ Refs.~\cite{Rosvall2014NatComm,Palla2007Nature,Mucha2010Science}. This view focuses on the individual nodes and seeks to group them in some principled way. If one, on the other hand, focuses on the communities instead of the nodes and prioritizes definitions that give interpretable communities (one temporal-network community could, for example, represent one seminar, one concert, etc.), it makes sense not to require every node to be a member of a group at every point in time~\cite{sune_fundamental}, as in Chapter 2.

Other mesoscopic structures, such as core-periphery structures~\cite{rombach2014core}, have been less studied for temporal networks (even though there are some works---\textit{e.g.}~\cite{rico2012abiotic} uses core-periphery analysis to understand ant-plant networks). Finally, \emph{temporally connected components} (see Chapter 6 and \cite{kivela_mapping_2018}) span the structural scale from mesoscopic to macroscopic, both in terms of network structure and with respect to time.

\subsection{Fundamental structures}

Chapter 2 and Ref.~\cite{sune_fundamental}, Lehmann \textit{et al.}\ discuss the traces that the six fundamental interaction types leave on temporal networks. Chapter 2 presents a division of the interaction types in the configuration of participants (one-to-one, one-to-many and many-to-many) and synchronicity (synchronous and asynchronous). In the limit of a short time interval projection of a temporal network data set, these different communication events contribute with different subgraphs---synchronous one-to-many communication yields a star graph, and synchronous many-to-many communication yields a clique. By tuning the time window, one can identify the time-scales of influence of these ``fundamental structures''.

\section{Important nodes, links and events}

Perhaps the most common question for static networks is to find important nodes (where ``important'' should be interpreted in an inclusive sense). This question is just as relevant for temporal networks. This is maybe the topic where the approaches borrowed from the static-network toolbox are most applicable to temporal networks. One major difference is that it is meaningful to talk about the importance of contacts (in addition to nodes and links) for temporal networks~\cite{Takaguchi_2012}. Another difference is that the most principled, general measures of importance are time-dependent, simply because, in most contexts, a node can become more or less important in time.

\subsection{Generalizing centrality measures}

A huge number of papers have been devoted to the generalization of classical centrality measures to temporal networks, see Refs.~\cite{Pan2011PhysRevE,taylor2017eigenvector} and Chapters 8 and 17. In many cases---for distance-based centrality measures---they have taken the obvious approach of replacing distances by latency. Since temporal networks are typically less connected (in the sense that the fraction of nodes that are reachable through time-respecting paths is smaller than the corresponding quantity in static networks), centrality measures have to work in fragmented networks. This means that one needs to combine information about how many nodes can be reached with information on how easily they can be reached (or whatever rationale the corresponding static centrality measure has). One example would be to generalize closeness centrality by averaging reciprocal latencies, rather than taking the inverse of the averages~\cite{thebook:latora}. This is, however, an arbitrary combination of two aspects of centrality and quite typical for straightforward generalizations of static concepts to temporal networks---they become less principled than their static counterparts.

\subsection{Controllability}

The rationales of centrality measures come from reasoning about dynamic systems---you can reach other nodes quickly from central nodes; central nodes are in the middle of heavy traffic, etc. The purpose of measuring centrality is typically to rank the nodes and perhaps list the most central ones. Finding \textit{control nodes} involves a slightly different thinking. Instead of ranking the nodes, the control nodes are minimal sets of nodes needed to be manipulated for the entire network to reach a certain state. Ref.~\cite{LiCornelius2017Science} and subsequent works show that temporal networks can facilitate controllability---\textit{i.e.}\ the system can be controlled with less energy and by fewer nodes if it has temporal heterogeneities.

\subsection{Vaccination, sentinel surveillance, and influence maximization}

The problems of vaccination, sentinel surveillance, and influence maximization are related to questions about spreading phenomena. Similarly to controllability, one assumes some objective and some intervention to the underlying temporal-network structure. In this case, however, the objective is typically to minimize or maximize the number of nodes reached by some spreading dynamics (like an infectious disease, word-of-mouth marketing, etc.).

The \textit{vaccination} problem is to select nodes that would minimize or slow down disease spreading as much as possible. Typically one assumes that the vaccinated nodes are deleted from the system so that they can no longer become infected and spread the disease. Unlike centrality measures, but similarly to controllability, it usually makes no sense to talk about the importance of individual nodes for the vaccination procedure---vaccinating one or a few nodes in a large network would have no measurable effect on the epidemics. Instead, the importance of nodes comes from the membership of a group that is vaccinated~\cite{jain}. Another important point is that one can typically not assume knowledge about the entire network of contacts---only the interactions that individuals could reliably report can serve as input for vaccination protocols. For example, Refs.~\cite{genois:vacc,starnini2013immunization} propose vaccination protocols that exploit temporal structures.

The \textit{influence maximization} problem deals with finding seed nodes for spreading dynamics that maximize the number of reached nodes~\cite{kempe2003maximizing}. The prime application is viral marketing, but to protect against outbreaks that have not yet entered a population influence maximization is also interesting for network epidemiology. The nodes that are important for vaccination and influence maximization do not necessarily have to be the same---optimal node sets for vaccination typically fragment the network efficiently. In contrast, influence maximization emphasizes efficiently splitting the network into subnetworks of influence. The first problem is akin to \textit{network attack} or \textit{network dismantling}~\cite{dismantling}, and the second to finding a \textit{vertex cover}~\cite{cover}. To exploit temporal structures, one can identify nodes in a heightened state of activity or nodes that reliably influence others (Chapter 18).

\textit{Sentinel surveillance} assumes that one can put sensors (sentinels) on the nodes. The task is to choose the sensors' locations such that disease outbreaks are discovered as reliably or quickly as possible. This is probably the least studied of these three problems on temporal networks---we are only aware of Ref.~\cite{bai2017optimizing}. On the other hand, it is practically a more important problem since it is currently in use in health care systems~\cite{sentinels}. Ref.~\cite{bai2017optimizing} tests how efficient temporal network protocols originally developed for vaccination are for the problem of sentinel surveillance.

\subsection{Robustness to failure and attack}

A problem that is very much overlapping with influence maximization etc., is network robustness. The scenario is that some adversary is trying to destroy a network. This adversary can have different amounts of information or resources to carry out the attack, which yields different versions of the problem. With no information about the network, the problem reduces to \textit{node percolation} (or \textit{robustness to failure}). With perfect information but limited computational resources, the problem is equivalent to network dismantling. It is both interesting to study optimal heuristics for this problem and what network structures contribute to the robustness of a network. For temporal networks, Refs.~\cite{tarjanovski,scellato_robu} studied this problem. Still, there should be several ways of extending their work, and in general, temporal-network robustness seems to be an understudied area. This may have to do with the fact that the temporal dimension makes the whole percolation framework more complicated (see Chapter 6).
 
\section{How structure affects dynamics}

For models of disease spreading, heterogeneous, heavy-tailed degree distributions are known to speed up the dynamics~\cite{Barthelemy2004PhysRevLett}. One line of research in  temporal network studies is to identify similar relations between the structure of the data and dynamics taking place over the contacts.

The types of dynamics people have been studying on underlying temporal networks include disease spreading of different kinds (Chapters 7, 11, 13, 14 and 16)~\cite{fffng}, threshold models of complex contagion~\cite{Takaguchi2013PlosOne,backlund}, random walks (Chapter 12)~\cite{starnini2012,delvenne_diffusion_2015,Masuda2016book,greedy}, navigation processes~\cite{lee:navigating}, synchronization (Chapter 15) and even game-theoretic models~\cite{cho:game,zhang:gaming}.

\subsection{Simulating disease spreading}

Disease spreading typically follows standard compartmental models developed by applied mathematicians~\cite{hethcote,BRITTON201024}. Such models divide a population into classes with respect to the disease, and then state transition rules between the classes. The key transition rule in all compartmental models is the \textit{contagion event} where a susceptible individual becomes infected when in contact with an infectious individual. In the two canonical and most well-studied models---the SIS (susceptible--infectious--susceptible) and SIR (susceptible--infectious--recovered) models---the contagion event is paired with the recovery of individuals (in SIS, recovered individuals become susceptible again, whereas in SIR they become immune to the disease or die). The probability that a contact between a susceptible individual and an infectious individual results in contagion is usually a model parameter. It is assumed to be the same for all contacts (which is an assumption for convenience and not realism).

Many assumptions are needed for simulating a compartmental model on a temporal network of contacts~\cite{masuda_holme_rev,ENRIGHT201888}. Unless modeling bio-terrorism or the spread of something other than a disease makes no sense to select more than one seed. Without prior knowledge about the disease's entry into the population, one should choose this seed uniformly at random. By the same principle, one should choose the time of infection uniformly randomly as well. This could, of course, lead to the outbreak not being able to reach all nodes, so that the measured outbreak sizes are an average of outbreak sizes of different times. For this reason, some authors choose to start the outbreak early in the interval their data covers, even though that introduces a bias if \textit{e.g.}\ the activity in the data grows~\cite{rocha_blondel}. Another commonly used approach is to use periodic boundary conditions and repeat the data from the beginning (\emph{e.g.}, in Ref.\ \cite{Karsai2011PhysRevE}).

Another important consideration is the duration of the infectious period. In the mathematical epidemiology literature, it is usually taken to be exponentially distributed to achieve the Markov property (that the probability of recovering is independent of the time since the infection). Markovian SIR and SIS are not only easier to analyze analytically but also allow for some tricks to speed up simulation code (see \url{github.com/pholme} for fast, event-driven code for the Markovian SIR on temporal networks). Some studies use a constant duration of infection for all nodes. To the best of our knowledge, no studies have tried duration distributions inferred from data.

Another decision that anyone simulating disease spreading (or random walks) on temporal networks needs to make is what to do with contacts happening in the same time step. There are, as we see it, two principled solutions. Either one assumes that this allowed, in which case one then needs to pick contacts with the same timestamp in random order and average over different randomizations, or one assumes the disease cannot spread via an intermediate node in a single time step. This is effectively to assume an SEIS or SEIR model (E stands for \textit{exposed}, which means that the individual will become infectious in the future but is not yet infectious), with the duration of the E state being less than the time resolution of the temporal network.

Another slight difference in approaches, especially in studies where a model generates the underlying temporal network, is that of \textit{link-centric} and \textit{node-centric} compartmental models. In node-centric models~\cite{masuda_rocha,non_poi}, the time to the next contact that could spread the disease is determined at a contagion event. In link-centric models~\cite{vaz,Horvath2014} the contacts are generated independently of the propagation of the disease. The node-centric model simplifies analytical calculations, whereas the link-centric model is conceptually simpler and perhaps more realistic (even though the assumption that the contact dynamics is independent of what spreads on the network is probably often invalid).

Typically, papers about disease spreading have focused on understanding how network structure affects the final outbreak size~\cite{Min2011PhysRevE,Holme2014SciRep,masuda_holme_rev}. Some, however, have studied early outbreak characteristics such at the basic reproductive number $R_0$ (the expected number of others an infectious individual would infect in a completely susceptible population)~\cite{liu_controlling,rocha_blondel}. From a medical perspective, there is no obvious choice between these two---even though the societal concern is to minimize the outbreak size. The outbreak size is also a consequence of interventions not modeled by canonical disease-spreading models such as SIS and SIR. Thus the early phase of the disease, which is better summarized by $R_0$, could be more informative.

Random-walk studies have focused on the \textit{mean-first passage time}---the expected time since the beginning of a walk that the walker reaches a node---and \textit{reachability}---the probability that a node is reached by a walker starting at a random node~\cite{Masuda2016book,greedy}. Another topic of interest has been how topological and temporal structure affects the speed of diffusion~\cite{delvenne_diffusion_2015}.
As opposed to the spreading of disease, there is no directly obvious real-world phenomenon that would be well-modeled by random walks on temporal networks; however, random walks equal diffusion, and diffusion can be considered a fundamental process in any system. Often, the random walk process is simply used as a probe of the temporal network structure.

\begin{figure}
\begin{center}
\includegraphics[width=\linewidth]{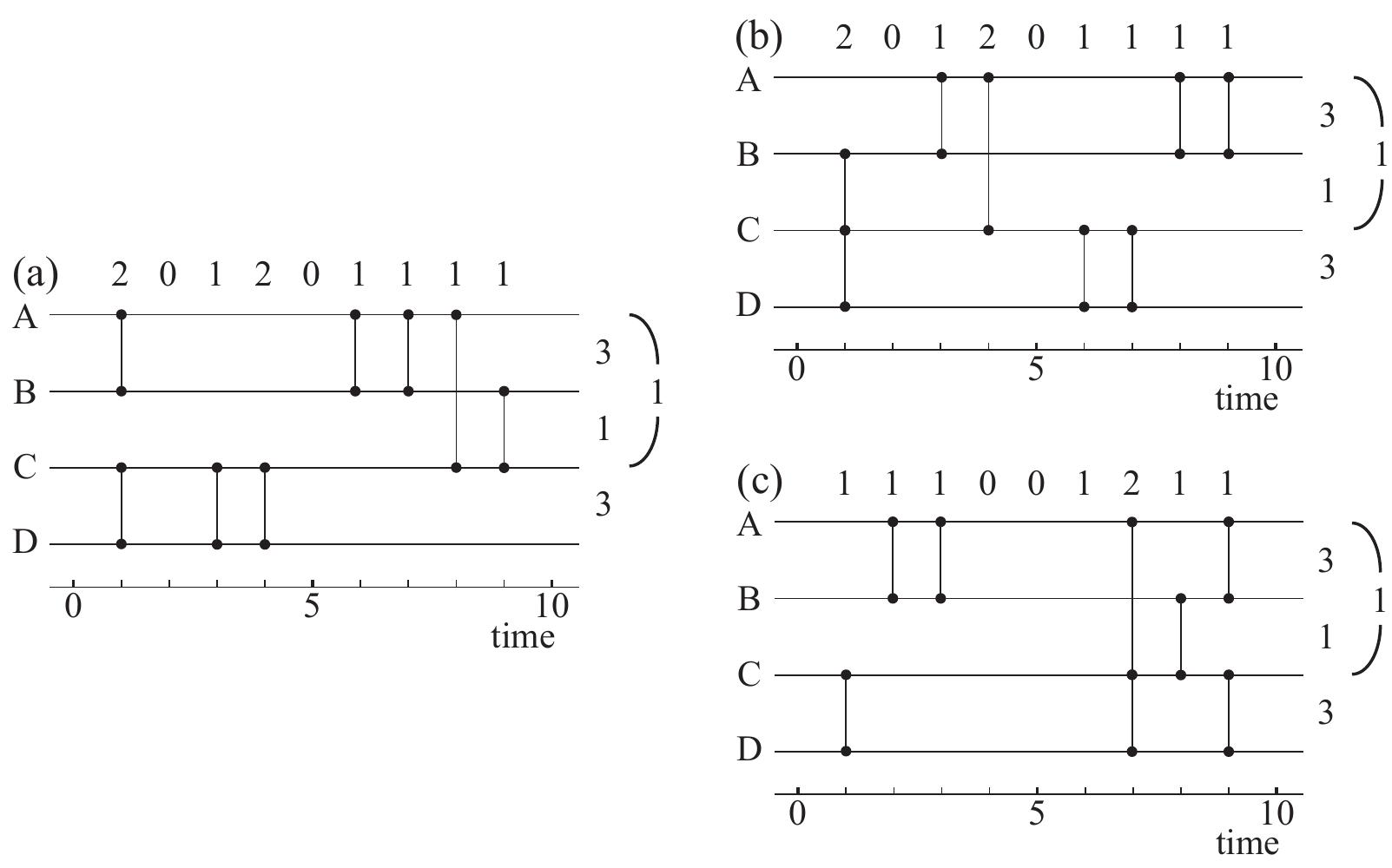}
\end{center}
\caption{Illustrating two types of randomization procedures. Panel (a) shows a temporal network that is randomized by randomly swapping time stamps (b) and replacing timestamps with random ones (c). The randomization in (b) preserves both the number of contacts per time (the numbers above) and the number of contacts per pair of nodes (the numbers to the right). The randomization procedure of panel (c) preserves the number of contacts per pair of nodes but not the number of contacts per time.
}
\label{fig:randomization}
\end{figure}

\subsection{Tuning temporal network structure by randomization}

The most straightforward way of understanding the impact of temporal network structure on dynamic processes is, of course, to tune it and monitor the response on some quantity describing the dynamics. There were important contributions (also involving temporal structures) in this direction even before the turn-of-the-millennium network boom. For example, Morris and Kretzschmar studied the effect of concurrency, or overlapping relations, on outbreak sizes~\cite{morris_kretzschmar}.

The most common way of investigating the effect of structures on a temporal network is to use randomization schemes. This approach starts with empirical networks and then destroys some specific correlation by randomizing it away. For example, one can randomly swap the timestamps of contacts or replace the timestamps with a random timestamp chosen uniformly between the first and last of the original data~\cite{holme2005}. The former randomization is more restrictive in that it preserves the overall activity pattern and per-node and per-link inter-event time statistics (see Fig.~\ref{fig:randomization}). Randomization schemes turn out to be much more versatile for temporal networks than for static ones. Ref.~\cite{gauvin} gives a comprehensive theory of almost 40 different randomization schemes. By applying increasingly restrictive methods to real data sets, one can see how much structure is needed to recreate the original temporal network's behavior.

In general, the terminology of temporal networks is ambiguous. The topic itself sometimes goes under the the names ``dynamic networks'', ``temporal graphs'', or ``time-varying networks''. The randomization schemes above are no exception---Ref.~\cite{holme2005} calls the scheme of Fig.~\ref{fig:randomization}(b) ``permuted times'', Ref.~\cite{Karsai2011PhysRevE} calls it ``shuffled times'' and Ref.~\cite{gauvin} calls it ``shuffled timetamps''.

\subsection{Models of temporal networks}

Another way of tuning temporal network structure, other than randomization, is by generative models. Generative models of temporal networks serve a  different role than static networks. Static network science traditionally used network evolution models as proof-of-concept models for theories about emergent properties, like power-law degree distributions~\cite{ba_model} or community structure~\cite{seceder}. There are common structures for temporal networks combining temporal and network structures in a non-trivial way that is non-trivial to explain. Nevertheless, models of temporal networks are needed, if for nothing else then to generate underlying data sets for controlled experimentation. In this section, we will mention some of the central developments in this area. For a complete overview, see Ref.~\cite{Holme2015EurPhysJB}.

\begin{figure}
\begin{center}
\includegraphics[width=0.8\linewidth]{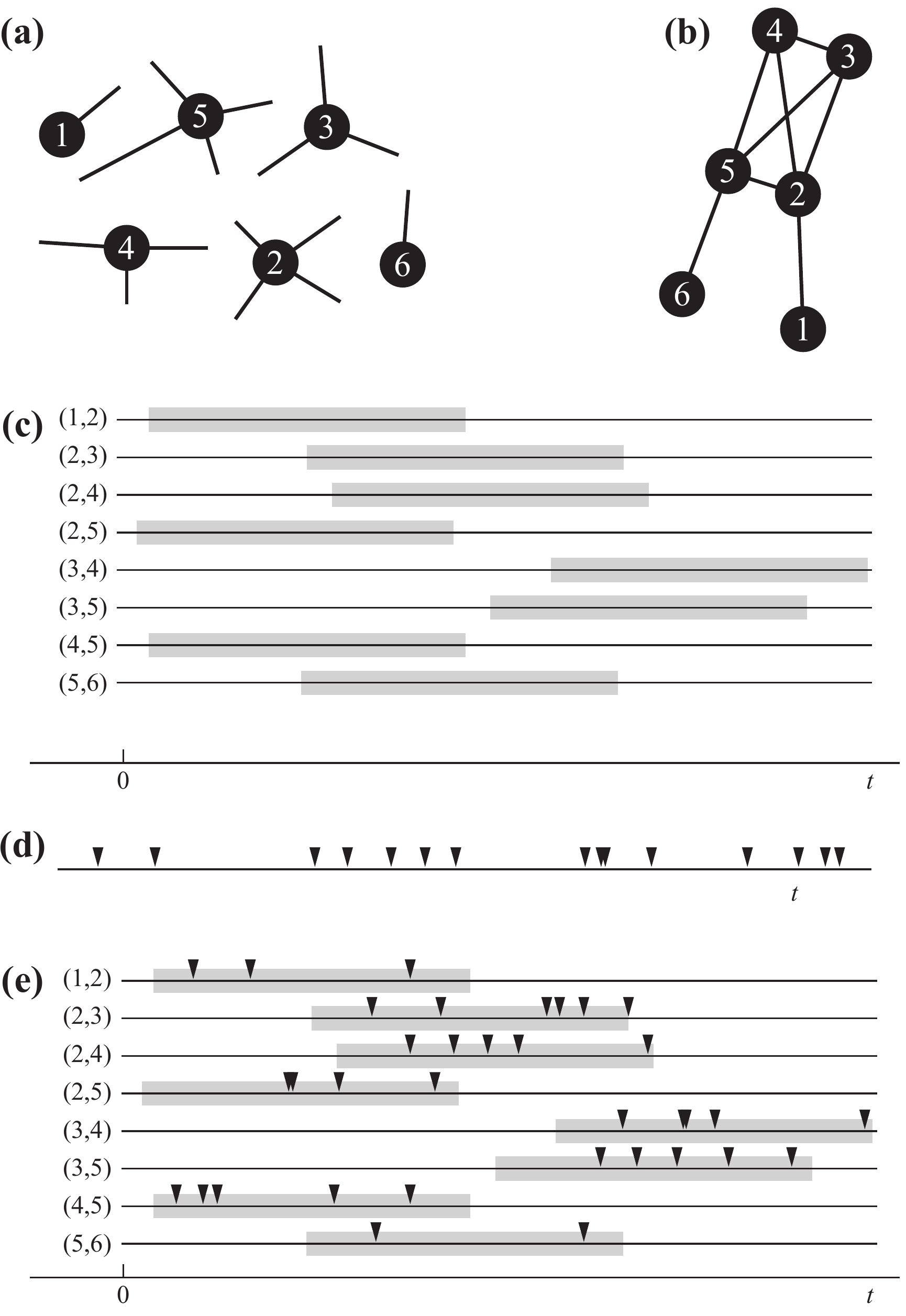}
\end{center}
\caption{Illustrating a simple generative model for temporal networks, used in Refs.~\cite{holme_comp_biol} and \cite{rocha_blondel}. First, one generates a static network from the configuration model by creating desired degrees from a probability distribution (a) and matching them up in random pairs (b). Then, one generates intervals for the links showing when they are active (c). Finally, one generates a time series of interevent times (d) and matches it to the active intervals. This figure is adapted from Ref.~\cite{Holme2015EurPhysJB}.
}
\label{fig:model}
\end{figure}

The most straightforward approach to generate a temporal network is to generate a static network and assign a sequence of contacts to every link. For example, Ref.~\cite{holme_comp_biol} uses the following procedure:
\begin{enumerate}
\item Construct a simple graph by first generating an instance of the configuration model~\cite{newman_book} and merging multiple links and self-links from it.
\item For every link, randomly generate an  interval when contacts can happen.
\item Generate a sequence of contacts following an interevent time distribution.
\item Match time sequence of contacts to the active intervals of the links.
\end{enumerate}
This model is illustrated in Fig.~\ref{fig:model}.

Perra et al.~\cite{perra_activitydriven} proposed a model of temporal networks---the \textit{activity driven model}---that is even simpler than the above with the advantage that it is analytically tractable. Let $G_t$ denote a simple graph at time $t$. Their generation algorithm proceeds as follows:
\begin{enumerate}
\item Increment the time to $t$ and let $G_t$ be empty.
\item Activate a node $i$ with probability $a_i\Delta t$. Connect $i$ to $m$ other randomly chosen distinct nodes.
\end{enumerate}
This model has been fundamental to analytical studies of processes on temporal networks, see \textit{e.g.}\ Refs.~\cite{perra_rw,karsai_perra,liu_metapop,liu_controlling,starnini_pastor,issue:perra,han_sun}.

Starnini et al.~\cite{pastor_face2face} uses a two-dimensional random walk model where the chance of approaching node $i$ is proportional to an increasing attractiveness parameter $a_i$. This means that the more attracted a walker is to its neighbors, the slower its walk becomes (simulating acquaintances stopping to talk when they meet on the street). Furthermore, they also allow people not to socialize by having occasional inactive periods. Zhang et al.~\cite{li_b} propose a similar model without an explicit representation of space.

Another model of temporal networks of social contacts was proposed in Ref.~\cite{vestergaard_how}. The authors introduced a model where temporal effects can activate both nodes and links. In their model, a link can be active or inactive and further characterized by the time $\tau_{(i,j)}$ since the last time it changed state. Similarly, node $i$ uses the time $\tau_i$ since it last was involved in a contact as a basis for its decisions. The network is initialized to $N$ nodes, and all links inactive. A node can activate a link with probability depending on $\tau$. The link is chosen from the nodes $i$ that are currently not in contact with $i$ with a probability depending on the $\tau$s of the neighbors. An active link is inactivated with a rate that is also dependent on $\tau$.

Refs.~\cite{thebook:masuda} and \cite{cho} use a Hawkes process to model a similar situation to the one considered by Starnini et al.~\cite{pastor_face2face} above. Ref.~\cite{thebook:masuda} argue that there is a positive correlation between consecutive interevent times in empirical data that one cannot model by interevent times alone. Their model works by defining an event rate by
\begin{equation} \label{eq:hawkes}
v+\sum_{i:t_i\leq t}\varphi(t-t_i)
\end{equation}
where $\phi$ is an exponentially decreasing memory kernel, and $v$ is a base event rate. Even with an exponentially decaying kernel, the interevent time distribution becomes broader than exponential. 
Similar to Refs.~\cite{thebook:masuda} and \cite{cho}, Ref.~\cite{colman_greetham} introduced a model of temporal networks based on stochastic point processes. In their model, nodes form and break links following a Bernoulli process with memory. Like the Hawkes process mentioned above, the probability of an event between $i$ and $j$ increases with the number of recent events between $i$ and $j$. Specifically, Ref.~\cite{colman_greetham} takes the probability of a link to activate or deactivate at time $t$ to be proportional to the number of such events in a time window.

\section{Other topics}

There are, of course, some themes in the temporal network literature that do not fit into the above three categories. Two examples are generalizations of \textit{link prediction}~\cite{linkprediction} and \textit{network reconstruction}~\cite{newman2018estimating,tiago_nwkrcnstr} to temporal networks. The motivation of both these topics is that real data is often erroneous and incomplete. In static networks, link prediction refers to the problem of finding the linkless pair of nodes that is most likely to be a false negative (falsely not having a link). In the context of temporal networks, this could be reformulated as either the question of what will be the next contact (given the information up to a point), or which contact was missing in the past. We are not aware of any paper addressing these particular problems. Instead of solving these purely temporal network questions, there is a large body of literature on link prediction in static networks with a turnover of nodes and links---see \textit{e.g.}\ Ref.~\cite{AHMED2016120} and references therein---\emph{i.e.},\  assuming a slower changing network that elsewhere in this chapter.

Network reconstruction, in general, is the problem of inferring a network from secondary, incomplete, or noisy data~\cite{newman2018estimating,tiago_nwkrcnstr}. So far, we are not aware of such temporal-network studies similar to the static network case. There are papers about the technical difficulties of inferring temporal network contacts from electronic sensors~\cite{Stopczynski,thebook:barrat} and papers about how to reconstruct static networks from temporal network data~\cite{window,holme_comp_biol}, but we are aware of no papers that would predict false positive and negative data in a contact sequence.

\section{Future perspectives}
Temporal network studies have been a vivid sub-discipline of network science for around a decade. Some issues of the early days have been settled, while others remain. This period has seen a shift from research that simply extends static-network ideas to temporal networks to methods that are unique to temporal networks. Still, the overall research directions are more or less the same as for static networks (\textit{cf.}\ Fig.~\ref{fig:triangle})---questions about identifying important nodes, how to simplify temporal networks further, and how their structure affects dynamics. Are there such larger research directions that make sense for temporal networks but not static ones? An obvious idea would be to focus on questions that involve time more directly. Only rarely have researchers asked what the optimal time to do something is, or the optimal duration to expose the system to some treatment, etc. Change-point detection (finding the time when a system changes between qualitatively different states) is one exception~\cite{peel2015detecting}. There are also papers about time series analysis of temporal networks~\cite{sikdar2016time,Huang_2017}, but these typically do not ask questions about time like the ones above.

Perhaps the crudest assumption of temporal network modeling to date (as mentioned in Sec.~\ref{sec:events}) is that the existence of a contact is independent of the dynamic system of interest. As an example, there are many modeling studies of information spreading on top of empirical temporal networks (\textit{e.g.}\ mobile-phone or e-mail data~\cite{Karsai2011PhysRevE,KARIMI,backlund}). Of course, information spreading via e-mails or calls does really happen. Still, one cannot usually view it as a random process on top of some temporal contact structure independent of the information. While one can imagine less important information spreading this way---``By the way, put on that Finnish heavy metal when uncle Fredrik comes to visit, he will appreciate it''---usually, calls are made and e-mails are sent with the explicit purpose of spreading information. Therefore, information spreading influences or even drives the contact structure. How should one then model information spreading on temporal networks? One possibility would be to give up using empirical data as the basis for the analysis; such an approach would be similar to \textit{adaptive networks}~\cite{GrossSayama2009book}. One could also go for data that contains the content of the messages or conversations instead of only their metadata; in this case, it might be possible to understand the relationship between contacts' temporal network and the spreading dynamics. Evidently, such data is hard to come by for privacy reasons, but interestingly, early studies of electronic communications did analyze both the content and the structure of spreading~\cite{crisis}. There are also communication channels where everything is public, such as Twitter.

One research direction with plenty of room for improvement is temporal-network visualization. Fig.~\ref{fig:connectivity} illustrates some of the challenges where Fig.~\ref{fig:connectivity}(a) gives a reasonable feeling for the temporal structures but none for the network structure, and for Fig.~\ref{fig:connectivity}(b), the situation is reversed. One can probably rule out a type of visualization that manages to show all information and convey all different aspects of the structure. However, there should be methods that discard some information but still reveal important structures. Also, animated visualization (that has the obvious limitation that not all the information is shown at once) probably has room for improvement. Some such methods are discussed in Chapter 5. The ``alluvial diagrams'' of Ref.~\cite{rosvall_alluvial} are another interesting approach. Evidently, there are some available methods, but we wish for an even wider selection to choose from.

Yet another fundamental challenge for temporal networks is how to rescale or subsample a data set properly. In particular, many methods, inspired by statistical physics, rely on ways to change the size of a network consistently. This is a challenge even for static networks---simply making subgraphs based on a random set of nodes will most likely change the structure of a network (other than Erd\H{o}s-R\'enyi random graphs)~\cite{sh}. The same goes for more elaborate ways of reducing the size of a network by merging nodes~\cite{bjk,song2006origins}---there is no guarantee that such manipulation will retain the structure of networks. For temporal networks, one might think that at least the temporal dimension could be rescaled by sampling windows of different sizes, but this is not trivial either because it could change whether or not a dynamic process has the time to reach a certain state or not. For finite-size scaling, such as used in the study of critical phenomena~\cite{hong2007finite}, one would need a way to link the size of the network and the duration of the temporal network.

Finally, as mentioned earlier in this chapter, we feel that there is a lot to do regarding temporal-network robustness and fragility, with applications ranging from network security to public health and the efficient planning of robust public-transport systems. This is an area where it is possible to go beyond static-network analogies. For example, while a static network may fragment when chosen nodes are attacked/immunized, the range of temporal-network responses is much broader. The network may remain temporally connected in principle, but the average latency of time-respecting paths may grow high enough to make them useless. Or, the system's latency could temporarily grow to make it temporarily disconnected: consider, \emph{e.g.}, congestion in a public transport system. Furthermore, the range of possible attack or immunization strategies can be much broader too: interventions to events, attacks that aim to increase latency generally, interventions at specific times, sequences of timed interventions at different nodes or contacts, and so on. Likewise, when the aim is to improve network robustness, interventions are not limited to network topology alone. For example, for improving the reliability of public transport systems, one could only modify the temporal sequences of connections and their time-domain statistics to minimize the disruption caused by random deviations from the planned schedules, or one could aim at maximal synchronization of connections to minimize the latency of time-respecting paths.

\begin{acknowledgement}
PH was supported by JSPS KAKENHI Grant Number JP 18H01655. JS acknowledges support from the Academy of Finland, project ``Digital Daily Rhythms'' (project no.\ 297195).
\end{acknowledgement}

\bibliographystyle{abbrv}
\bibliography{bib}

\end{document}